\documentclass{article}[11pt]
\newcommand{\code}[1]{\texttt{#1}}
\newcommand{\baseURL}{\code{baseURL} }
\usepackage{graphicx}
\usepackage{hyperref}
\begin{document}

\title{The OAI Data-Provider Registration and Validation Service}
\author{Simeon Warner\\
Cornell Information Science, Ithaca, NY 14850, USA,\\
\code{simeon@cs.cornell.edu},\\
\url{http://www.cs.cornell.edu/people/simeon}}

\date{$ $Date: 2005/06/03 15:57:28 $ $}
\maketitle

\begin{abstract}
I present a summary of recent use of the Open Archives Initiative (OAI)
registration and validation services for data-providers. The registration
service has seen a steady stream of registrations since its launch in 2002, 
and there are now over 220 registered repositories. I examine the 
validation logs to produce a breakdown of reasons why repositories fail 
validation. This breakdown highlights some common problems and will be 
used to guide work to improve the validation service.
\end{abstract}

\section{Introduction}

The Open Archives Initiative (OAI) released the OAI Protocol for Metadata
Harvesting (OAI-PMH) in 2001, to facilitate metadata harvesting from data-providers
(repositories). A validation service was launched coincident with the 
initial protocol release to allow data-providers to check compliance with 
the protocol, and has been updated for versions 1.1 and 2.0 of the 
OAI-PMH~\cite{OAI-PMHv2}.

In 2001 there were no standard OAI libraries or repository packages implementing 
the protocol, so every deployment of the OAI-PMH had new code to be tested. 
Since then several libraries and software packages implementing the protocol 
have become available and these have eased compliance problems. However, in 2003
an OAI harvesting project reported that over 10\% of repositories had XML
errors~\cite{HALBERT+03}. The validation service has helped identify errors 
in popular software packages (e.g. DSpace and eprints.org) and in particular
deployments of these and other packages. Several other facilities are also 
available to test OAI-PMH implementations, the most important of which is
the Repository Explorer~\cite{SULEMAN01}.

In this paper I present a brief analysis of registrations (section 2) and 
validation requests (section 3) received via the 
OAI website\footnote{\url{http://www.openarchives.org/Register/ValidateSite}}
during 2004. I then (section 4) discuss these results in the context of new 
work to improve the validation facilities.

\section{Registration}

A key function of the OAI data-provider validation facility is to build and maintain a 
centralized list of OAI-PMH compliant repositories\footnote{{OAI} registration service, 
list of registered repositories: \url{http://www.openarchives.org/Register/BrowseSites}}.
Registration is a voluntary way to announce the availability of a data-provider and
the registry has been a useful starting point for harvesting projects. 
It is now supplemented by additional registries including those of
Celestial\footnote{Celestial OAI Registry: \url{http://celestial.eprints.org/cgi-bin/status}},
eprints.org\footnote{eprints.org registry of Institutional Repositories: \url{http://archives.eprints.org/eprints.php}} and
OLAC\footnote{{OLAC} archives registry: \url{http://www.language-archives.org/archives.php4}},
and even what might be thought of as a `virtual registry' through Google search~\cite{HABING03}.

\begin{figure}[ht]
\begin{center}
\includegraphics[scale=0.9]{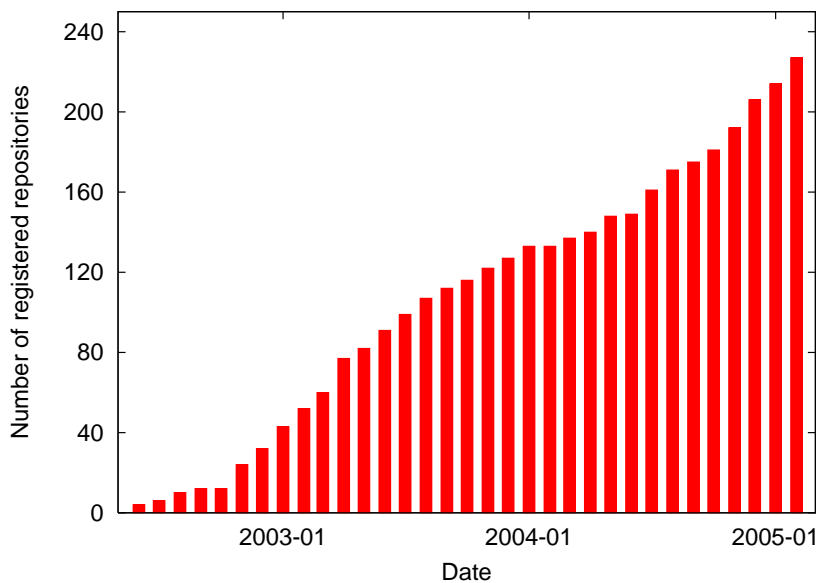}
\end{center}
\caption{\label{fig-registered-dps} Number of registered OAI-PMH v2.0 data-providers 
as a function of time since the release of v2.0 in June 2002.}
\end{figure}

Figure~\ref{fig-registered-dps} shows the number of data-providers, or repositories, 
registered with the OAI validation service as a function of time since version 2.0 of 
the protocol was released in June 2002 (see~\cite{LAGOZE+03} for data covering the 
period from 2001--2002 with earlier protocol versions). Earlier versions of the 
protocol were deprecated on release of version 2.0, and the 86 registrations of version 1.0 
and 1.1 repositories were discarded in December 2002. Figure~\ref{fig-registered-dps} 
shows that in just a few months the number of version 2.0 registrations exceeded that 
for earlier versions, indicating an effective transition. The steady increase in the 
number of registered sites suggests that the registration facility continues to 
valued by the community.

\section{Validation}

Registration requests differ from validation requests only in that,
if validation is successful, registration requests result in the 
\baseURL\footnote{For details of the \baseURL see: 
\url{http://www.openarchives.org/OAI/2.0/openarchivesprotocol.htm\#HTTPRequestFormat}}
being entered in the central registry. In the following I consider all 
requests together and refer to them as `validation requests'.
Table~\ref{tab-requests} shows a total of 1893 validation requests 
logged during 2004. An OAI-PMH \baseURL could 
be extracted, and thus validation attempted, from 95\% of requests. 
In the remaining 5\% of cases, the web form was either not filled-in or 
an invalid \baseURL was given, and thus no further tests could be made.

\begin{table}[ht]
\begin{center}
\begin{tabular}{|l|r|r|}
\hline
                                         &     Number &   \% total \\
\hline
                             No base URL &         89 &        4.7 \\
                       Nonsense base URL &          7 &        0.4 \\
                          Valid base URL &       1797 &       94.9 \\
\hline
                         Total requests  &       1893 &      100.0 \\
\hline
\end{tabular}%
\end{center}
\caption{\label{tab-requests} Validation requests logged during 2004.}
\end{table}

There are a number of error conditions which cause the validator to abort
validation tests. These conditions include fundamental errors such as the
wrong protocol version being reported in the \code{Identify} response,
and errors where it is not possible to extract data required for subsequent tests.
Table~\ref{tab-fatals} shows a breakdown of the reasons for aborted validation requests. 
In 40\% of aborted validations there was no response to the \code{Identify} request, 
usually because the \baseURL was entered incorrectly (user error or an interface 
issue, not a problem with the protocol). 
In 21\% of cases, bad XML was returned resulting in failure to parse 
the \code{Identify} response. Here the validator returns the diagnostic output from the 
Xerces\footnote{Xerces {XML} parser and validator: \url{http://xml.apache.org/\#xerces}}
XML validator. While this output is specific about both the location
and reason for an error, it is rather difficult to interpret without detailed knowledge of 
the W3C XML schema specification and the particular schema being used. In many cases such
errors were corrected quickly though in a significant minority additional explanation
and/or help was requested via email.
As the response to the \code{Identify} request is particularly important, several other 
checks are made on this response and bad protocol version and bad administrator email
address errors are highlighted in the table. 
Errors in the response to the \code{Identify} request are often the result of incomplete 
repository setup or simple administrator mistakes. In most cases they were quickly corrected 
in response to validator error messages.

\begin{table}[ht]
\begin{center}
\begin{tabular}{|l|r|r|}
\hline
                                         &     Number &   \% total \\
\hline
                    No Identify response &        349 &       40.1 \\
\hline
       Failed to parse Identify response &        184 &       21.1 \\
             Bad protocol version number &          3 &        0.3 \\
                 Bad admin email address &         62 &        7.1 \\
     Other errors with Identify response &        212 &       24.4 \\
\hline
Excessive 503 \code{Retry-After} replies &          9 &        1.0 \\
     No identifiers from ListIdentifiers &         29 &        3.3 \\
           No datestamp in sample record &         22 &        2.5 \\
\hline
                  Total aborted requests &        870 &      100.0 \\
\hline
\end{tabular}
\end{center}
\caption{\label{tab-fatals} Breakdown of aborted validation requests by reason.}
\end{table}
 
The last three reasons shown in table~\ref{tab-fatals} are more indicative of
problems with the repository implementations. The validator correctly handles HTTP 
503 \code{Retry-After}\footnote{\url{http://www.openarchives.org/OAI/2.0/openarchivesprotocol.htm\#HTTPResponseFormat}} responses by waiting the specified period and then retrying. 
However, in some cases repositories repeatedly give \code{Retry-After} responses 
and 1\% of the aborted validations were because of more than 5 successive \code{Retry-After} responses.
The last two reasons relate to tests that obtain the identifier of a sample item 
from a \code{ListIdentifiers} response. In 3.3\% of cases there were no items in the 
repository and validation was aborted because it is not possible to comprehensively 
test an empty repository. In a further 2.5\% of cases, no datestamp could be extracted 
from the sample record. Without the datestamp of a sample record it is not possible to
test datestamp-based incremental harvesting requests. This error indicates a mistake
in the implementation of OAI-PMH records as all records must have a datestamp.

\begin{table}[ht]
\begin{center}
\begin{tabular}{|l|r|r|}
\hline
Error                           &  Number \\
\hline
Schema validation errors in standard verb responses  & 168 \\
Empty response when \code{from} and \code{until} set to known \code{datestamp} & 57 \\
Empty \code{resumptionToken} in response to request without \code{resumptionToken} & 42 \\
Malformed response to request with identifier \code{invalid"id} & 40 \\ 
Granularity of \code{earliestDatestamp} doesn't match \code{granularity} value & 35 \\
\hline
\end{tabular}
\end{center}
\caption{\label{tab-poperrors} Most common validation errors in cases where validation
was completed.}
\end{table}

Of the 927 completed validation requests, 318 were successful, 198 had errors only in the 
handling of exception conditions and 411 had other errors. Failures occurred
in all conditions tested although certain failures were particularly common. 
The 5 most common errors are shown in table~\ref{tab-poperrors}.

\begin{figure}[ht]
\begin{center}
\includegraphics[scale=0.9]{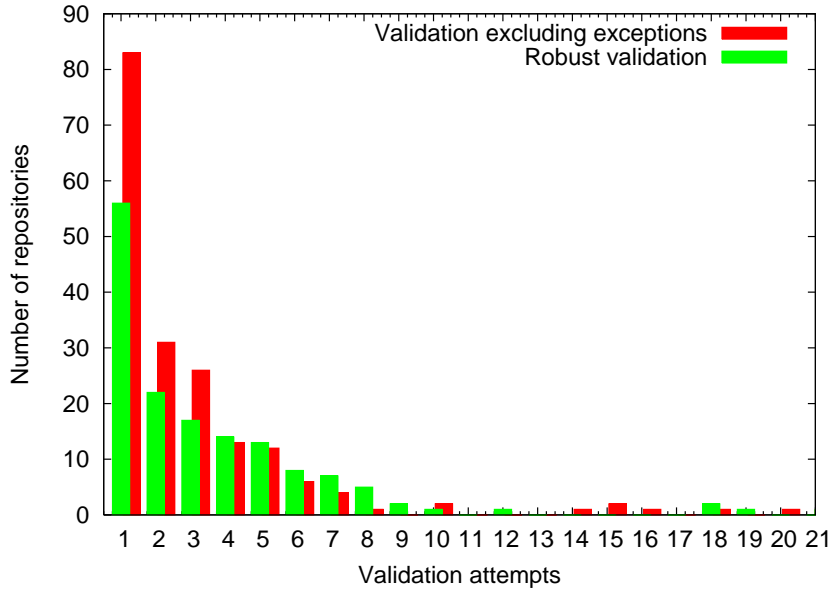}
\end{center}
\vspace*{-5mm}
\caption{\label{fig-tries} Histogram of attempts before successful validation. Two cases are shown: 
`validation excluding exceptions' indicates that the repository passed all tests using 
valid requests; `robust validation' indicates that the repository not only passed all 
tests using valid requests but also correctly responded with appropriate exception and
error codes to various illegal requests. In cases where repositories failed validation after
initially passing, only the number of validation attempts until the first success are 
counted.}
\end{figure}

To show how many attempts were required to validate each data-provider, 
figure~\ref{fig-tries} histograms validation attempts per repository before 
success. The figure separates cases where repositories pass all tests using valid 
requests (`validation excluding exceptions') from cases where repositories also 
correctly respond to various illegal requests (`robust validation').
46 of the 152 robustly validating repositories achieved validation excluding exceptions before 
robust validation, and it took about 3 further attempts on average to correct problems in
the responses to illegal requests. 33 repositories managed validation excluding exceptions but
never passed robust validation and were thus not eligible to register.
In most cases only a few validation attempts were required before success, and in 38\% of cases, 
often deployments of standard software, validation was successful on 
the first attempt. Not shown in the graph are 376 sites which never passed validation. Of 
these, 238 can be discounted as validation was attempted only once, typically trial runs
and test deployments. 
Most tried just a few times though a significant tail of 24 repositories failed validation 
more than 5 times, which suggests a serious attempt to validate, yet were never successful. 
These cases were investigated and table~\ref{tab-failure6} shows a breakdown of the current
status of these repositories.

\begin{table}[ht]
\begin{center}
\begin{tabular}{|l|r|r|}
\hline
Reason                           &  Number \\
\hline
Repository no longer accessible  &  14 \\  
Repository reports OAI-PMH v1.1  &   1 \\
Internal server errors (HTTP 500) &  2 \\
Old DSpace with OAI problem      &   1 \\
XML errors                       &   1 \\
Repository can be harvested from successfully &   5  \\
\hline
Total number of repositories examined &  24 \\
\hline
\end{tabular}
\end{center}
\caption{\label{tab-failure6} Breakdown of current status of repositories that
failed validation tests more than 5 times and we never successful.}
\end{table}

The 5 repositories shown in table~\ref{tab-failure6} as being able to be harvested
successfully all still fail the validation test. One simply returns server errors. 
One includes style-sheet information in the XML responses which Xerces cannot 
parse. The other three all fail under certain exception conditions. All three
incorrectly handle a \code{GetRecord} request for metadata from an item with
the illegal identifier \code{invalid"id} by attempting to include that value as
an attribute of the \code{request} element \textit{and} by failing to escape the
quotation mark.

\section{Discussion and future work}

The continued use of the validation facility and the registration of new repositories
attests to the value of these services. It is reassuring to see that most repositories
manage to correct errors and pass the validation test in just a few attempts. However, 
personal assistance has been provided in a number of cases and a significant number of 
sites tried to validate several times but never succeeded. These cases suggest 
that there is room for improvement in the protocol documentation and the helpfulness of 
the validation suite.

Improvements in the validation facility should include more detailed explanation of 
common error conditions. There should also be specific tests that help identify common
XML errors which we see are still common. Once identified it should be possible
to provide messages that are more helpful than the standard Xerces output.
A number of frequently recurring errors such as correctly dealing with an illegal
identifier should be easy to correct. However, there is some subtlety in the 
specification and it has often required off-line email exchange to clarify the
issue. More detailed on-line explanations of common or confusing cases may
help address this.

The results presented here show a number of cases where repositories work 
sufficiently well to be harvested from yet fail strict compliance tests. Perhaps
it is time to re-evaluate the decision to provide only black and white, registration
or failure. The registration site might be augmented with a status that could indicate,
for example, basic compliance (valid requests work), robust compliance (exception
conditions also handled correctly), and compliance with Dublin Core (to allow sites 
that don't use the \code{oai\_dc} Dublin Core metadata to check compliance with the rest of 
the protocol).

The analysis presented here is the first step in a project to produce improved OAI 
validation tools for the NSDL\footnote{http://nsdl.org/} and the broader OAI community. 
Future work will include refinement of the existing validation suite, and development of 
validation and testing software for harvesters through the development of test repositories 
displaying various error conditions. New facilities will be announced to the OAI community 
through the usual email list\footnote{OAI-implementers email discussion list and archive: 
\url{http://www.openarchives.org/mailman/listinfo/oai-implementers}}
and on the OAI website\footnote{http://www.openarchives.org/}.

\section{Acknowledgements}

The software used in this work was a direct development of the 
validation software written by Donna Bergmark between 2001 and 2003. 
I thank Naomi Dushay for pointing out a number of validation issues. 
This work is supported by NSF grant number 0127308.

\bibliographystyle{plainurl-sw2003-09-29}
\bibliography{dl}

\end{document}